\RequirePackage{ifpdf}
\ifpdf % We are running pdfTeX in pdf mode
\documentclass[pdftex]{sigma}
\else
\documentclass{sigma}
\fi

\numberwithin{equation}{section}

\begin{document}

\allowdisplaybreaks

\renewcommand{\thefootnote}{$\star$}

\renewcommand{\PaperNumber}{047}

\FirstPageHeading

\ShortArticleName{Translation-Invariant Noncommutative Renormalization}

\ArticleName{Translation-Invariant Noncommutative\\ Renormalization\footnote{This paper is a
contribution to the Special Issue ``Noncommutative Spaces and Fields''. The
full collection is available at
\href{http://www.emis.de/journals/SIGMA/noncommutative.html}{http://www.emis.de/journals/SIGMA/noncommutative.html}}}

\Author{Adrian TANASA~$^{\dag\ddag}$}

\AuthorNameForHeading{A.~Tanasa}

\Address{$^\dag$~Centre de Physique Th\'eorique, CNRS, UMR 7644,\\
\hphantom{$^\dag$}~\'Ecole Polytechnique, 91128 Palaiseau, France}
\EmailD{\href{mailto:adrian.tanasa@ens-lyon.org}{adrian.tanasa@ens-lyon.org}}

\Address{$^\ddag$~Institutul de Fizic\u a \c si Inginerie Nuclear\u a Horia Hulubei,\\
\hphantom{$^\ddag$}~P.O.\ Box MG-6, 077125 M\u agurele, Rom\^ania}

\ArticleDates{Received March 25, 2010, in f\/inal form May 24, 2010;  Published online June 08, 2010}

\Abstract{We review here the construction of a translation-invariant scalar model which was proved to be perturbatively renormalizable on Moyal space. Some general considerations on nonlocal renormalizability are given. Finally, we present perspectives for generalizing these quantum f\/ield theoretical techniques to group f\/ield theory, a new setting for quantum gravity.}

\Keywords{noncommutative quantum f\/ield theory; Moyal space; locality; translation-inva\-riance}

\Classification{81T18; 81T75}

\renewcommand{\thefootnote}{\arabic{footnote}}
\setcounter{footnote}{0}

\section{Introduction}

Noncommutative geometry (see for the example the book \cite{book-cm}) can play some role in fundamental physics.
When considering f\/ield theory on noncommutative Moyal space, the Grosse--Wulkenhaar model \cite{GW} was a f\/irst proposition for a renormalizable scalar model.
Let us also mention the absence of the Landau ghost in this model \cite{beta-GW} (in contrast to commutative $\phi^4$ theory).
Nevertheless, the Grosse--Wulkenhaar model has an important drawback: it is not translation-invariant.

Recently, a translation-invariant scalar model was proposed and also proved renormalizable on Moyal space \cite{GMRT}. The main idea is to consider the model modif\/ied by its own quantum corrections (which are translation-invariant, unlike the Grosse--Wulkenhaar harmonic oscillator term).

On a general basis, let us emphasize that noncommutative quantum f\/ield theories (for a~ge\-ne\-ral review, see for example
\cite{Szabo}) can  be interpreted as limits
of matrix models or of string theory models. The f\/irst use of
noncommutative geometry in string theory was in the formulation of
open string theory  \cite{witten}. Noncommutativity is here natural
just because an open string has two ends and an interaction which
involves two strings joining at their end points shares all the
formal similarities to noncommutative matrix multiplication. In this
context, one also has the Seiberg--Witten map \cite{sw}, which maps
the noncommutative vector potential to a conventional Yang--Mills
vector potential, explicitly exhibiting the equivalence between
these two classes of theories.

But probably  the simplest context in which noncommutativity  arises is in a~limit in which a~large background
antisymmetric tensor potential dominates the background metric. In
this limit, the world-volume theories of Dirichlet branes become
noncommutative \cite{string1,string2}. Noncommutativity was
also recently proved to arise as some limit of loop quantum gravity
models. There, the ef\/fective dynamics of matter f\/ields coupled to
$3$-dimensional quantum gravity is described after integration over
the gravitational degrees of freedom by some noncommutative quantum
f\/ield theory~\cite{fl}.
In a dif\/ferent context, some $3$-dimensional noncommutative space emerging in the context of  $3$-dimensional Euclidean quantum gravity was also studied in \cite{karim}.

In condensed matter physics, noncommutative theories can be of particular interest
when describing ef\/fective nonlocal interactions, as is the case, for example, of the fractional
quantum Hall ef\/fect. Dif\/ferent authors proposed that a good description of this
phenomenon can be obtained using noncommutative rank~1 Chern--Simons theory \cite{hall1}.

This review presents the translation-invariant model \cite{GMRT} as well as some general ideas behind this nonlocal renormalizability. The review is structured as follows. In the next section
the mathematical setting, namely the Moyal space, is presented. The third section exposes the issue of the UV/IR mixing. The fourth section presents the translation-invariant model and some general considerations on the associated Feynman rules and perturbative renormalizability.
The f\/ifth section shows some explicit calculation of Feynman integrals. A rough sketch of the renormalization proof and some details on a possible mechanism for taking the commutative limit are also given.
Some further f\/ield theoretical implementations are then brief\/ly listed in the sixth section.
The last section presents perspectives for generalizing these nonlocal techniques for quantum group f\/ield theory, a new attempt for writing a quantum theory of gravitation.

\section[Mathematical framework - the Moyal space; implementation of field theory]{Mathematical framework -- the Moyal space;\\ implementation of f\/ield theory}

The Moyal algebra is the linear space of smooth and rapidly decreasing functions ${\cal S}({\mathbb R}^D)$ equipped with the {\it Moyal product}
\begin{gather}
\label{produs}
(\phi \star \psi)(x)=\int \frac{d^D k}{(2\pi)^D} d^Dy \, \phi\left(x+\frac 12 \Theta \cdot k\right) \psi(x+y) e^{ik\cdot y}, \qquad \forall\, \phi,\psi,
\end{gather}
where the noncommutative matrix $\Theta$ has the block-diagonal form
\begin{gather*}
%\label{theta}
\Theta  =  \begin{pmatrix} \Theta_2 & 0 & \ldots  \\
&  \vdots  &  & \\
0 & \ldots & \Theta_2 \end{pmatrix},\qquad
\Theta_2  =  \begin{pmatrix} 0 & -\theta \\ \theta & 0 \end{pmatrix}. \nonumber
\end{gather*}

Let us remark that this product is manifestly nonlocal, noncommutative but associative. Taking the limit $\theta\to 0$, the integral on $k$ in the def\/inition \eqref{produs} leads to a $\delta$-distribution of $y$ and one obtains the commutative product of the respective functions.

{\sloppy
One can extend by duality (considering the product of a tempered distribution with a~Schwartz-class function) the algebra def\/ined above to contain amongst others, the identity, the polynomials, the $\delta$-distribution and its derivatives. In this larger space, the following identity holds:
\[
[x_\mu, x_\nu]_\star=x_\mu\star x_\nu-x_\nu\star x_\mu=i\Theta_{\mu\nu}.
\]
This relation is sometimes given as the def\/ining relation of the Moyal space.

}

When implementing f\/ield theories on such a noncommutative space, the most common way is to simply replace the usual multiplication of f\/ields by the noncommutative Moyal multiplication \eqref{produs}. Nevertheless, let us emphasize that this is equivalent to the implementation of noncommutativity through the use of ordinary products in the noncommutative ${\mathbb C}^\star$-algebra of Weyl operators (see for example \cite{Szabo}).

To end this section we brief\/ly address the question of the choice of Moyal noncommutative geometry. This particular product has the advantage of being a simple one (given by a constant parameter $\theta$, a deformation of the product of the commutative ${\mathbb C}^\star$-algebra of functions on space-time). Involved calculation can thus be explicitly performed. Moreover, when considering noncommutativity as a limit of string theory in a large background antisymmetric tensor, it is a Moyal-like geometry that is obtained.

The action for the Euclidean $\phi^4$ model on the Moyal space thus writes (in position space):
\begin{gather}
\label{act-x}
S[\phi]=\int d^4 x \left[ \frac 12 \partial_\mu \phi \star \partial^\mu \phi
+\frac 12 m^2 \phi^{\star\, 2}+\frac{\lambda}{4!}\phi^{\star\, 4}\right].
\end{gather}
In momentum space, this action becomes
\begin{gather*}
%\label{act-p}
S[\phi]=\int d^4 p \left[ \frac 12 p_\mu \phi \star p^\mu \phi
+\frac 12 m^2 \phi^{\star\, 2}+\frac{\lambda}{4!}V^\star[\phi]\right],
\end{gather*}
where $V[\phi]$ is the corresponding potential.

Let us now investigate what the consequences are of the use of the Moyal product \eqref{produs} in the action above. At the level of the kinetic part, using the def\/inition \eqref{produs}, one obtains the following general identity:
\[
\int d^4 x (\phi\star \psi)(x)=\int d^4 x \phi (x) \psi (x), \qquad \forall \,  \phi, \psi.
\]
This means that the propagation of the theory \eqref{act-x} is not af\/fected by noncommutativity: it remains identical to the one of the corresponding commutative model.
This af\/f\/irmation is true on an Euclidean space (which is the case treated here).
Let us note that, when considering Minkowskian theories with noncommutative time, the concept of time-ordering has to be gene\-ra\-li\-zed (see for example~\cite{Sibold}).

Nevertheless, at the level of the interaction term, things change in a crucial manner. Thus, again through a direct use of the def\/inition \eqref{produs} of the Moyal product, one can prove
 \begin{gather}
\label{para}
\int d^4 x \phi^{\star\, 4}(x)= \int \prod_{i=1}^4 d^4 x_i \phi (x_i)\delta(x_1-x_2+x_3-x_4)e^{i(x_1-x_2)\wedge(x_3-x_4)},
\end{gather}
where $x\wedge y = 2 x\Theta^{-1}y$.

One thus has a manifestly nonlocal, parallelogram-shaped interaction. The oscillation factor above is proportional to the area of the respective parallelogram.
Another important consequence of the utilization of the Moyal product is that, if in a commutative model the interaction is symmetric under permutation of the incoming/outgoing f\/ields, now the interaction is symmetric only under {\it cyclical} permutation.

For the sake of completeness, we end this section by stating that some notions of
noncommutative $\varepsilon$-graded connections were introduced and applied to Moyal space in \cite{axel0}.

\section{The UV/IR mixing}

When doing perturbation theory, the action \eqref{act-x} leads to the appearance of a new type of divergence~-- the UV/IR mixing \cite{min}.
\begin{figure}[t]
\centerline{\includegraphics[width=8cm]{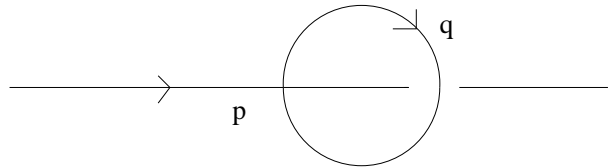}}
\caption{A tadpole graph in noncommutative quantum f\/ield theory.}
\label{np-tad}
\end{figure}
 This can be understood when computing the planar irregular\footnote{We denote by irregular the graphs for which the external legs ``break'' more then just one of the faces of the graph,
see for example~\cite{GMRT} for details.}
 tadpole (see Fig.~\ref{np-tad}) amplitude:
\begin{gather}
\label{np-tad-ampli}
\lambda \int d^4 q \frac{e^{\imath q_\mu \Theta^{\mu\nu}p_\nu}}{q^2+m^2}.
\end{gather}
Because of the presence of the Moyal oscillating factor, this amplitude is now convergent. It computes to
\[
\frac{\lambda}{4\pi^2}\sqrt{\frac{m^2}{(\Theta\cdot p)^2}}K_1 \Big(\sqrt{m^2 (\Theta\cdot p)^2}\Big).
\]
In the IR regime of the momentum $p$, this behaves like
\begin{gather*}
%\label{behaviour}
\frac{1}{\theta^2p^2}.
\end{gather*}
The same type of behavior was proved at {\it any order in perturbation theory} in \cite{limita}, for a planar irregular $2$-point graph.

When inserting such a graph into some ``bigger'' graph, the resulting graph will be nonplanar. The external momentum $p$ becomes internal and one has to integrate on. This leads to the appearance of a divergence in the IR regime of~$p$~-- the UV/IR mixing. This name is a~translation of the fact that one has an IR divergence in $p$ after one has integrated over the UV regime of $k$.

Such a divergence is manifestly nonlocal and it appears at the level of the $2$-point function. If the theory was renormalizable, it should be absorbed in a redef\/inition of the parameters of the quadratic part of the action (i.e.\ the mass or the wave function). But, as already stated above, the quadratic part of the action is local and such a redef\/inition cannot be achieved; this leads to the non-renormalizability of the model \eqref{act-x}! Furthermore, this is not a peculiarity of scalar models, but rather a general feature for f\/ield theories on Moyal space (see again \cite{min}).

The same type of divergence appears for quantum f\/ield theories implemented with other type of translation-invariant $\star$-products as well (like for example the Wick--Voros product) \cite{pat}.

Let us end this section by stating that the issue of UV/IR mixing
has also been studied  from a dif\/ferent point of view,
in terms of (emergent) gravity, see for example \cite{stein}.

\section{Translation-invariant renormalizable model;\\ consideration on renormalizability}

\subsection{The model; Feynman rules}

As already stated in the Introduction, the f\/irst model proved renormalizable on Moyal space was the Grosse--Wulkenhaar model \cite{GW}, whose propagator is modif\/ied by a harmonic oscillator term. This additional term restores at the level of the whole action the symmetry between position and momentum~-- the Langmann--Szabo invariance \cite{LS}:
\begin{gather}
\label{har}
\Delta S_{\rm GW} [\Phi (x)] = \int d^4 x \,\frac 12   \frac{\Omega^2}{\theta^2}  x^2  \Phi^2.
\end{gather}
This formalism can be extended to scalar f\/ield theories def\/ined on noncommutative Minkowski space \cite{szabo2}.

Symmetries of the Euclidean theory above were extensively investigated in \cite{axel}. The classical action is invariant under the orthogonal group if this group acts also on the symplectic structure. It was found that the invariance under the orthogonal group can be restored also at the quantum level by restricting the symplectic structures to a particular orbit.

Another model of this type is the Langmann--Szabo--Zarembo model \cite{LSZ}, a model of a noncommutative scalar in an external background magnetic f\/ield.
Several f\/ield theoretical or algeb\-raic geometrical tools have then been implemented for these types of model (see \cite{dimreg,fab,param2,param,param',mellin,marcolli,goldstone} and references within).

As already stated above, a dif\/ferent proposition for a renormalizable model  was recently made in \cite{GMRT}.
The main idea behind this proposition is to write an action where the propagation is modif\/ied to take into account its own quantum corrections; thus, the action writes (in momentum space)
\begin{gather}
\label{revolutie}
S_\theta [\phi]=\int d^4 p \left(\frac 12 p_{\mu} \phi  p^\mu \phi  +\frac
12 m^2  \phi  \phi
+ \frac 12 a  \frac{1}{\theta^2 p^2} \phi  \phi
+ \frac{\lambda }{4!} V^\star\right ).
\end{gather}
where  $a $ is  some dimensionless parameter, which is free to f\/low under renormalization group (just as the parameter $\Omega$ in \eqref{har} was).
The propagator is
\begin{gather}
\label{propa}
C(p,m,\theta)=\frac{1}{p^2+m^2+\frac{a}{\theta^2 p^2}}.
\end{gather}

Let us also specify that the vertex contribution in an arbitrary Feynman amplitude is the usual Moyal one; in momentum space this writes:
\begin{gather*}
%\label{faza-vertex}
\delta(p_1+\cdots +p_4)e^{-\imath\sum_{i<j}p_i^\mu\Theta_{\mu\nu}p_j^\nu}.
\end{gather*}
The $\delta$ function corresponds to the usual momentum conservation at the vertex (supposing here that all momenta are incoming), while the oscillating phase above is a pure consequence of the presence of noncommutativity. This formula represents the Fourier transform of the position-space vertex contribution \eqref{para}.

Let us end this subsection by emphasizing that the proposed model thus changes only the propagator but not the vertex contribution to an arbitrary Feynman amplitude. This characteristic is also present in the Grosse--Wulkenhaar model.

\subsection[Mixing of scales - key of noncommutative renormalization]{Mixing of scales -- key of noncommutative renormalization}

The key of noncommutative renormalization (in the Grosse--Wulkenhaar case or in the case of the model \eqref{revolutie}) is the mixing of scales.
The renormalization group scales are given by the inverse of the propagator. Thus, in a commutative theory, this inverse is simply $p^2$ and for high values of $p^2$ one speaks of the UV regime, while for small values of $p^2$ one speaks of the IR regime.

In renormalizable theories on Moyal space, these scales are not so easy to def\/ine because they are mixed. Thus, the inverse of the propagator~\eqref{propa} is $p^2+\frac{a}{\theta^2p^2}$ and this function is big for both big and small values of the momentum $p$.
The value of $p$ which minimizes this expression is
\[
p=\sqrt{\frac{a}{\theta^2}}.
\]
One can thus speak in this sense of some kind of mixing of the UV and IR scales, just as it happens when computing Feynman amplitudes (see the previous section).
Furthermore, this could be interpreted as some kind of RG f\/low from the extremum (big or small) to the ``medium''.

A similar phenomena happens also for the Grosse--Wulkenhaar model, where, as already stated  above, instead of the $1/p^2$ term one has some harmonic oscillator term. From the point of view of the scale decomposition discussed here, the two terms are similar, since $x^2$ is big/small for small/big value of $p$.

We thus conclude that, even if locality is lost when uplifting to a noncommutative setting, renormalizability can still be achieved. A new type of principle which underlies this, is the so-called principle of ``Moyality''. This can be understood by the fact that nonlocal (but however Moyal-like) counterterms (of the same form as the terms in the bare action) are required to cure the perturbative divergences

Let us give some further explanation of the principle of ``Moyality'' in position space.
 In commutative theories the divergences are local; this is consistent to the fact that the terms in the bare action are local as well~-- the physical principle of locality.
In Moyal f\/ield theories the divergences are nonlocal, but they appear however when the external positions form a~paralle\-lo\-gram. This is again consistent with the form of the bare interaction \eqref{para}. Thus the physical principle of locality described above is generalized in a natural manner to the one of ``Moyality''.

The interested reader may report himself to
\cite{beta-GMRT} or \cite{io-kreimer} for a more general discussion on this generalization.

For the sake of completeness, let us also state that comments on nonlocality versus Wilsonian renormalization were made in \cite{rosten}.

\section{Feynman calculations; insights on the renormalizability proof\\ and on the commutative limit}

In this section we give some calculational details, using the example of the planar irregular tadpole graph of Fig.~\ref{np-tad} and a more involved example which  illustrates the way in which the UV/IR mixing is cured.

Let us f\/irst notice that the propagator \eqref{propa}
 decomposes for $a<\theta^2 m^4/4$ as
\begin{gather*}
C(p,m,\theta)=\frac{1}{p^2+m^2}-\frac{1}{p^2+m^2}\frac{a}{\theta^2 p^2 (p^2+m^2)+a}
\nonumber\\
\phantom{C(p,m,\theta)}{}=
\frac{1}{p^2+m^2}-\frac{1}{p^2+m^2}\frac{a}{\theta^2 (p^2 +m_1^2)(p^2+m^2_2)},%\label{propa2}
\end{gather*}
where $-m_1^2$ and $-m_2^2$ are the roots of the denominator of the second term in the f\/irst line of the RHS, considered as a second order equation in $p^2$, namely
\[
\frac{-\theta^2 m^2\pm \sqrt{\theta^4 m^4 - 4 \theta^2 a}}{2\theta^2}<0.
\]
Thus the Feynman amplitude of the tadpole graph of Fig.~\ref{np-tad} writes
\begin{gather}
\label{np-tad-ampli-GMRT}
\lambda\int d^4 q {e^{\imath q_\mu \Theta^{\mu\nu}p_\nu}}\left(\frac{1}{q^2+m^2}-\frac{a}{\theta^2}\frac{1}{(q^2+m^2)\prod_{i=1}^2(q^2+m_i^2)}\right).
\end{gather}
The f\/irst term above is nothing but the one of \eqref{np-tad-ampli} thus leading to a behavior of type $\frac{1}{p^2}$. By a simple power counting argument, one can prove that the second term in \eqref{np-tad-ampli-GMRT} is convergent. Thus, this tadpole has the same type of divergence as in the case of the model \eqref{act-x}. Nevertheless, in the case of the theory \eqref{revolutie}, this divergence can now be {\it a priori} absorbed in a redef\/inition of the parameter $a$; this was not the case for the theory~\eqref{act-x}.
Note the same type of result can be obtained for any value of the parameter $a$ (see for example~\cite{GMRT}).

\begin{figure}[t]
\centerline{\includegraphics[width=7.5cm]{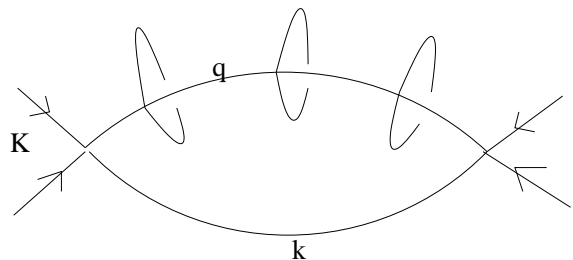}}
\caption{A nonplanar graph obtained from the insertions of several planar irregular tadpole graphs like the one of Fig.~\ref{np-tad}. In the model \eqref{act-x} this graph is IR divergent, while in the modif\/ied model \eqref{revolutie}, this graph is IR convergent, because of the presence of the $1/p^2$ term in the propagator.}
\label{N}
\end{figure}

Let us now investigate the behavior of the Feynman amplitude of the more general graph obtained from inserting a chain of $N$ planar irregular tadpoles in the $1$-loop $4$-point planar regular graph (see Fig.~\ref{N}).
We denote the internal momenta circulating the $N$ tadpole graphs by $p_i$, $i=1,\ldots, N$.
The integral to investigate writes
\begin{gather}
\label{unu}
\int d^4 q \prod_{i=1}^Nd^4p_i\frac{e^{\imath p_i^\mu \Theta_{\mu\nu}q^\nu}}{p_i^2+m^2+\frac{a}{\theta^2p_i^2}} \left(\frac{1}{q^2+m^2+\frac{a}{\theta^2q^2}}\right)^{N+1}
\frac{1}{(q-K)^2+m^2+\frac{a}{\theta^2(q-K)^2}}.
\end{gather}
Let us now have a closer look at the structure of the divergences of this general integral. When performing the integrations in the momenta $p_i$ ($i=1,\ldots,N$) and placing ourselves in the IR regime of the momentum $q$, each of these integrals leads to a $1/{(\theta^2 q^2)}$ behavior (as proved above). The integral \eqref{unu} thus becomes
\begin{gather*}
%\label{doi}
\int d^4 q
\left(\frac{1}{\theta^2q^2}\right)^N
\left(\frac{1}{q^2+m^2+\frac{a}{\theta^2q^2}}\right)^{N+1}
\frac{1}{(q-K)^2+m^2+\frac{a}{\theta^2(q-K)^2}}.
\end{gather*}
Note that  if $a=0$ this integral is IR divergent for $N>1$ (for $N=1$  the mass $m$ prevents the divergence to appear). Nevertheless, if $a\ne 0$,  in the IR regime of $q$ the dominant term is the $a/{\theta^2q^2}$ in the propagators and the integral
leads to an IR f\/inite behavior.

After these examples, let us now give a rough sketch of the renormalizability proof. The method used in \cite{GMRT} to prove this is the multi-scale analysis within the BPHZ renormalization scheme.

Let us recall here the following table summarizing the renormalization of the model  compared  to the ``one'' of the model \eqref{act-x} (which is not renormalizable):

\begin{center}
\begin{tabular}{|l|p{18mm}|p{18mm}|p{18mm}|p{18mm}|p{15mm}|p{15mm}|}\hline
& \multicolumn{2}{|c|}{{\it model \eqref{act-x}}}
& \multicolumn{2}{|c|}{{\it Grosse--Wulkenhaar model}}
& \multicolumn{2}{|c|}{{\it  model \eqref{revolutie}}}\\
\cline{2-7}
 & 2P & 4P & 2P & 4P & 2P & 4P\tsep{2pt}\\
\hline
planar reg & ren & ren & ren & ren & ren & ren\\
\hline
planar irreg & UV/IR & log UV/IR & conv & conv & f\/inite ren & conv \\
\hline
nonplanar & IR div & IR div & conv & conv & conv & conv \\
\hline
\end{tabular}
\end{center}
where ``ren'' means renormalizable, ``conv'' (resp.\ ``div'')  means convergent (resp. divergent), ``reg'' (resp. ``irreg'') means regular (resp.\ irregular).
Furthermore, ``2P'' and resp.~``4P'' mean $2$-point and resp. $4$-point Feynman graphs. We deal with them here because these are the graphs indicated to be primitively divergent by the power counting theorem proved in \cite{GMRT}. A similar power counting result was proved by a dif\/ferent method~-- the parametric representation~--  in~\cite{param-GMRT}.

Let us now present the main idea of the mechanism presented in \cite{limita} for obtaining the commutative limit of the model \eqref{revolutie}. Note that the new term in the action has a divergent, ``naive'' limit when $\theta\to 0$. This feature is also present in the Grosse--Wulkenhaar model.

The strength of the mechanism proposed in \cite{limita} comes from the fact that the new term is directly dictated by the quantum correction of the propagator (at any loop order, as proved in \cite{limita}). When letting $\theta\to 0$ in this type of Feynman amplitude, one obtains the usual wave function and mass renormalization of commutative $\phi^4$ theory. Hence, when $\theta$ is turned of\/f, the $1/(\theta^2 p^2)$ must not be present.
One splits the usual mass and wave function counterterms (in the commutative $\phi^4$ theory) into two parts. The f\/irst is again a mass and wave function counterterm corresponding to the planar irregular graphs (present when $\theta\to 0$) and the second part is the $a$ counterterm (present only when $\theta\ne 0$).
Taking all this into account, in \cite{limita} was proposed the following
action with ultraviolet cutof\/f $\Lambda$
\begin{gather*}
S_{\Lambda, \theta} [\phi] = \int d^4 p \left[\frac 12  \eta^{-1}(p/\Lambda) p^2 \phi^2 + \frac 12 m^2 \phi^2 +
\frac{\lambda}{4!} V_\theta
+ \frac 12 \delta_{Z'} p^2 \phi^2  + \frac 12 \delta_{Z''} (  1-T(\Lambda, \theta))
p^2 \phi^2  \right.  \nonumber\\
 \phantom{S_{\Lambda, \theta} [\phi] =}{}
 + \frac 12 \delta_{m'}\phi^2
 +\frac 12 \delta_{m''} \phi^2  (1-T(\Lambda, \theta))
+\frac 12 \delta_a \frac{1}{\theta^2 p^2} \phi^2 T(\Lambda, \theta)
+\frac 12 a \frac{1}{\theta^2 p^2} \phi^2 T(\Lambda, \theta) \nonumber\\
 \left.\phantom{S_{\Lambda, \theta} [\phi] =}{}
+\frac 12 \delta_{m'''} \phi^2  (1-T(\Lambda, \theta))
+\frac{\delta_{\lambda'}}{4!} (1-T(\Lambda, \theta))V_\theta
+\frac{\delta_{\lambda''}}{4!} T(\Lambda, \theta)V_\theta
\right],%\label{act-limita}
\end{gather*}
where we have written the counterterms associated to \eqref{revolutie}. The cutof\/f $\Lambda$ is some ultraviolet scale with the dimension of a momentum. The function $\eta^{-1}(p/\Lambda) $ is a standard momentum-space ultraviolet cutof\/f which truncates momenta higher than $\Lambda$ in the propagator (\ref{propa}). For that
$\eta(p)$ could be a f\/ixed smooth function with compact support interpolating smoothly
between value~1 for $\vert p/\mu \vert \le 1/2$ and 0 for $|p/\mu| \ge 1$.
Furthermore $T(\Lambda, \theta)$ is some smooth function satisfying the following conditions:
\begin{gather*}
\lim_{\theta \to 0} T(\Lambda, \theta)\frac{1}{\theta^2} =  0,\qquad
\lim_{\Lambda \to \infty} T(\Lambda, \theta) =  1.%\label{limite}
\end{gather*}
There are of course inf\/initely many functions which satisfy these conditions, for instance a~possibility is
\begin{gather*}
%\label{candidat}
T(\Lambda, \theta)= 1- e^{-\Lambda^6 \theta^3}
\end{gather*}
(where the factor in the exponential has been chosen to be dimensionless).
The interested reader may report himself to
\cite{limita} for  details.

\section{Further f\/ield theoretical developments}

The model presented in the previous sections is the f\/irst model on
Moyal space which is both renormalizable and translation-invariant. Moreover, the modif\/ied
propagator
\eqref{propa}
appears independently in recent work on non-Abelian gauge theory in the context of the
Gribov--Zwanziger result \cite{dudal}. Within a dif\/ferent framework, the same type of noncommutative propagator was studied from a phenomenological point of view in relation to the CMB \cite{patil}. The static potential in such a framework was also computed in \cite{coulomb}.

The beta-functions of the noncommutative model \eqref{revolutie} were
calculated in \cite{beta-GMRT}.
Moreover, as already stated above, the
parametric
representation of this model was implemented in \cite{param-GMRT}. Using this method, the explicit dependence of the superf\/icial degree of divergence on the graph genus can be computed.

The combinatorial properties of the polynomials of the parametric representation of the model~\eqref{revolutie} were extensively investigated in \cite{br}. A proof of the connection
between these polynomials and polynomials known by mathematicians in graph theory (the Bollob\'as--Riordan polynomial) was given.

The combinatorial aspects of the Dyson--Schwinger
equations within this noncommutative Moyal framework were also analyzed \cite{io-kreimer}. This equation is written as a series of powers
of some operators which inserts primitively divergent graphs into primitively divergent graphs.
These operators are, from a mathematical point of view, Hochschild $1$-cocycles on the Hopf algebra underlying Moyal renormalization.
Applying the renormalized Feynman rules to these combinatorial equations allows one to write
down, in a recursive manner, the Dyson--Schwinger equations used when doing non-perturbative
physics.
These combinatorial techniques  have then been gene\-ra\-li\-zed to the dif\/ferent framework of the spin-foam formalism
of loop quantum gra\-vi\-ty~\cite{io-sf}.

The same kind of recipe for curing the UV/IR mixing was then extended to f\/ield theories based on translation-invariant $\star$-products other then the Moyal product (like for example the Wick--Voros product) \cite{io-pat}.

The idea behind the model \eqref{revolutie} was extended to propose a modif\/ied $U(1)$ gauge model on the Moyal space \cite{gauge-GMRT}. This type of gauge theory has a trivial vacuum; this a crucial dif\/ference with respect to gauge theories based on a Grosse--Wulkenhaar-like modif\/ication, which were proved to have a highly nontrivial vacuum state \cite{gauge-GW}.

\section[Perspectives - from noncommutative field theory to quantum gravity (group field theory)]{Perspectives -- from noncommutative f\/ield theory\\ to quantum gravity (group f\/ield theory)}

An  important perspective of the quantum f\/ield theoretical approaches presented here is their extension for the study of the renormalizability properties of quantum gravity models.
The group f\/ield theory formalism of quantum gravity (see for example the reviews of~\cite{gft}) is the most adapted one for such a study, since it is formulated in a f\/ield theoretical setting.
These models were developed as a generalization of
$2$-dimensional matrix models to the $3$- and $4$-dimensional cases. Thus, group f\/ield theoretical
models are duals to the Ponzano-Regge model, when considering the $3$-dimensional gravity,
or to the Ooguri model, when considering the topological $4$-dimensional one.

The natural candidates for generalizations of matrix models in higher dimensions ($>2$)
are tensor models. The elementary cells that, by gluing
together form the space itself, are the $D$-simplices ($D$ being the dimension of space). Since a
$D$-simplex has $(D+1)$ facets on its boundary, the backbone of group f\/ield theoretical models in
$D$-dimension should be some abstract $\phi^{(D+1)}$ interaction on rank $D$ tensor f\/ields $\phi$.

Group f\/ield theoretical models can be seen today not only as a technical tool but as a~proposition for a quantum formulation of gravitation. Behind this lies the general idea that
group f\/ield theories are theories of space-time, while quantum f\/ield theories are theories on
space-time.
Let us also emphasize that the f\/ields of the models are written not as
functions of space-time (as is done in quantum f\/ield theories) but of group elements of some
non-Abelian group (like for example $SU(2)$ or $SO(4)$). If in the case of noncommutative f\/ield theories, the space-time itself is
noncommutative, in the case of group f\/ield theories noncommutativity is now given by the
non-Abelianity of the group one writes the model on. This group depends of course of the
dimensionality of the space but also on the Euclidean or Minkowskian signature one works
with.

\begin{figure}[t]
\centerline{\includegraphics{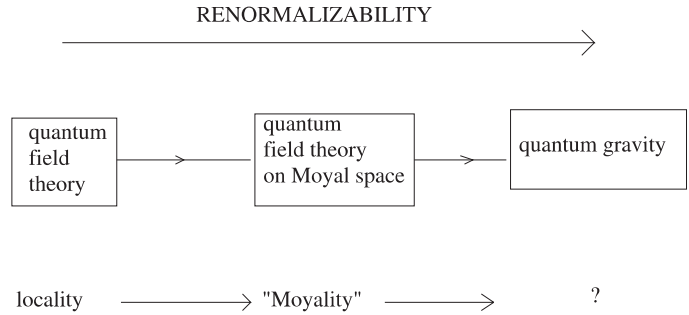}}
\caption{Renormalizability is used an an Ariadne's thread to guide us in the labyrinth of proposed physical models. In commutative f\/ield theory, renormalizability is based on the principle of locality. When going to f\/ield theories on the Moyal space, locality is lost. Nevertheless, it can be replaced by a~new principle, of ``Moyality''. The perspective suggested in this section is to investigate whether or not a generalization to some kind of principle of ``triangularity'' or ``simpliciality'' exists.}
\label{plan}
\end{figure}

The main perspectives are, as already stated above, related to the investigation of  the renormalizability properties
of dif\/ferent group f\/ield theoretical models proposed as candidates for a quantum theory of
gravity. This can be achieved by an appropriate generalization of the techniques
reviewed here for the study of the renormalizability of nonlocal theories. One can
investigate the generalization of the principle of ``Moyality'' to a new one, of ``triangularity''
or ``simpliciality'' (since one deals in these models with triangulation of space-time,
generalized then by the Ponzano--Regge and Ooguri models). This can be illustrating in Fig.~\ref{plan}.

A f\/irst such perspective is given by
the study of the renormalizability of topological group f\/ield theoretical models. Promising results have been obtained recently for $3$- and $4$-dimensional models \cite{ultim}. The next step is thus the generalization of these techniques to non-topological quantum gravity models, like for example the ones proposed, in a spin-foam formalism however, in \cite{eprl} and \cite{fk}.

\subsection*{Acknowledgements}

The author  was partially supported by the CNCSIS grant ``Idei'' 454/2009, ID-44 and by the grant PN 09 37 01 02.

\pdfbookmark[1]{References}{ref}
\LastPageEnding

\end{document}